# Aromaticity in Polyacene Analogues of Inorganic Ring Compounds

**Pratim Kumar Chattaraj* and Debesh Ranjan Roy**

*Department of Chemistry, Indian Institute of Technology, Kharagpur 721302, India.*

The aromaticity in the polyacene analogues of several inorganic ring compounds (BN-acenes, CN-acenes, BO-acenes and $Na_6$-acenes) is reported here for the first time. Conceptual density functional theory based reactivity descriptors and the nucleus independent chemical shift (NICS) values are used in this analysis.

In this communication we report for the first time the aromaticity in the polyacene analogues of several inorganic ring compounds (BN-acenes, CN-acenes, BO-acenes and $Na_6$-acenes). Conceptual density functional theory based reactivity descriptors and the nucleus independent chemical shift (NICS) values are used for this purpose.

Most popular benzene like inorganic aromatic compounds include s-triazine, borazine, boraxine, $Na_6$ etc. with planar $D_{3h}$ ($D_{6h}$) symmetry. Because of their simiar connectivity patterns[1] (Figures S1-S5) as that of benzene these molecules are expected to show aromaticity.[2] Several experiments authenticate their aromatic behavior albeit with some qualitative difference like more electron localization around the more electronegative atoms and preference of addition over substitution reactions. Aromaticity in the polyacene analogues of borazine is studied.[3] In this communication we will make use of various conceptual density functional theory (DFT)[4,5] based descriptors and associated electronic structure principles, viz., maximum hardness principle(MHP),[6] minimum polarizability principle(MPP),[7] minimum electrophilicity principle (MEP),[8] etc. as well as the nucleus independent chemical shift[9] values calculated at the ring center, NICS(0), and 1Å above the ring, NICS(1). The statements of these principles are as follows: a) MHP: "There seems to be a rule of nature that molecules arrange themselves so as to be as hard as possible", b) MPP: "The natural direction of evolution of any system is towards a state of minimum polarizability", and c) MEP: "The electrophilicity ($\omega$) will be a minimum (maximum) when both chemical potential ($\mu$) and hardness ($\eta$) are maxima (minima)".

Geometries are optimized at the B3LYP/6-311+G(d) level for $C_{4n+2}H_{2n+4}$, $(BN)_{2n+1}H_{2n+4}$, $(CN)_{2n+1}H_{n+2}$ : n=1-5; $(BO)_{2n+1}H_{n+2}$ : n=1,2. For polyacene analogues of boraxine with n=3-5 no convergence is archived. Single point calculations (B3LYP/6-311+G(d)) on the experimental geometries[10] of polyacene analogues of $Na_6$ are performed for the $Na_6$ units without the ligands. All the molecules are found to be planar (except $C_7N_7H_5$). Chemical potential[11]($\mu$), hardness[12]($\eta$) and electrophilicity[13]($\omega$) are calculated using the formulas:

$$\mu = -\frac{I+A}{2}, \ \eta = \frac{I-A}{2}, \ \omega = \frac{\mu^2}{2\eta}, \ I \text{ and } A \text{ being the}$$

ionization potential and electron affinity respectively, calculated using the koopmans' approximation in terms of highest occupied and lowest unoccupied molecular orbital energies.

The geometrical parameters, energies and other DFT descriptors are provided as the Supplementary Information (Figures S1-S5, Tables S1-S5). Figure 1 and Table 1 depict the NICS values. All the single ring (n=1) compounds ($C_6H_6$, $B_3N_3H_6$, $C_3N_3H_3$, $B_3O_3H_3$, and $Na_6$) reveal negative NICS(0) and NICS(1) values and hence are aromatic. Like polyacenes all the borazine analogues are aromatic but as shown earlier[3] the inner rings of borazine analogues are less aromatic than the corresponding outer rings unlike in polyacenes.[14] However, a perfect resemblance with the polyacene aromatic behavior is observed in the related $Na_6$ analogues. Most of the CN-acenes and BO-napthalene are antiaromatic.

Table 2 delineates that ionization potential (*I*), electron affinity (*A*), hardness ($\eta$), chemical potential ($\mu$), electrophilicity ($\omega$), energy (E), and polarizability ($\alpha$) can be expressed as quadratic functions of n (the number of rings in various polyacene analogues). Figure 2 confirms the linear behavior between $\alpha^{1/3}$ and 2S (1/$\eta$), as expected.[15]

An analysis of the behavior of energy, hardness and polarizability values per ring reveals that with an increase in n the energy/ring value increases whereas the hardness/ring value decreases as expected (Figures S6, S7) from the maximum hardness principle. The polarizability/ring also increases in most cases (except BN-acenes) as per the minimum polarizability principle. The minimum electrophilicity is not expected to hold good as the magnitude of the chemical potential does not increase monotonically with n.[8]

Aromatic/antiaromatic behavior of polyacenes, BN-acenes, CN-acenes, BO-acenes and $Na_6$-acenes are analyzed in terms of nucleus independent chemical shift and various conceptual DFT based reactivity descriptors. Most of the polyacene analogues of the inorganic ring compounds are aromatic in nature albeit with some qualitative differences in their aromatic behavior with that of polyacenes. Some of these inorganic ring compounds are antiaromatic..

**Acknowledgement.** We thank BRNS, Mumbai for financial assistance.

**Supporting Information Available:** Geometrical parameters, energies and other DFT descriptor values and their variations with n, of the polyacenes and their inorganic analogues. This material is available free of charge via the Internet.

* To whom correspondence should be addressed. E-mail: pkc@chem.iitkgp.ernet.in

FIGURES

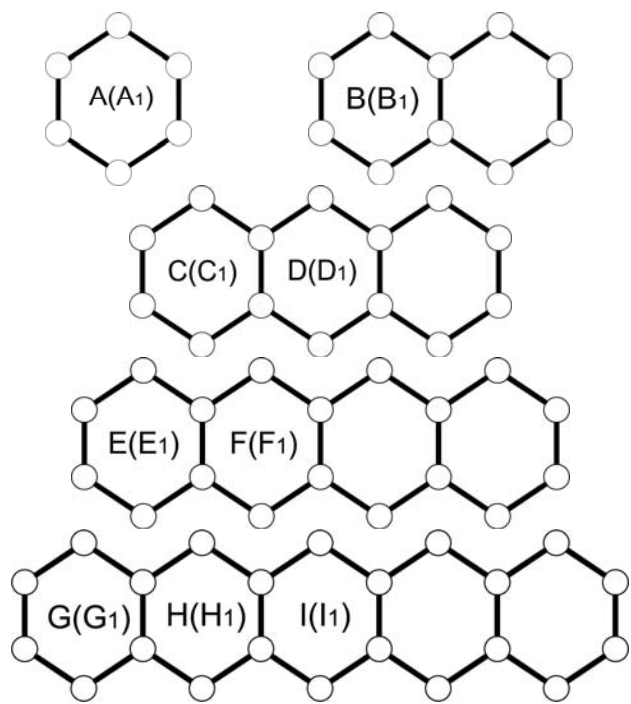

**Figure 1.** NICS(0)(NICS(1)) values at the ring center and 1Å above the plane respectively of the polyacenes.

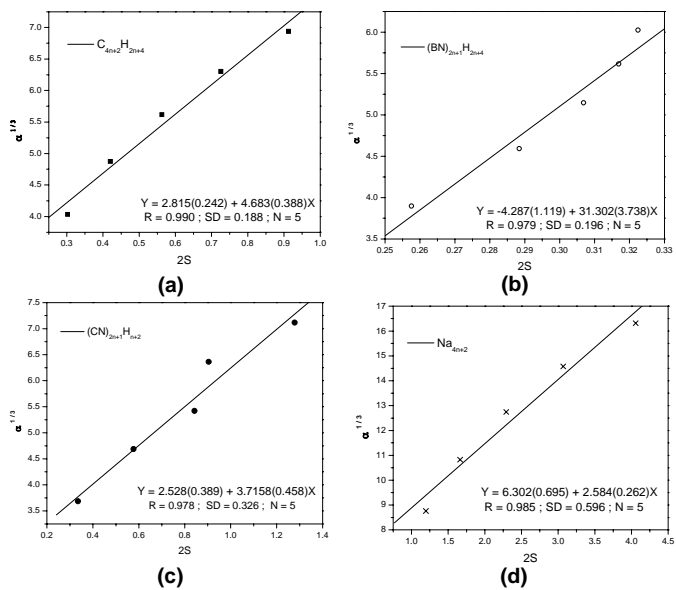

**Fig. 2.** $\alpha^{1/3}$ vs 2S of (a) $C_{4n+2}H_{2n+4}$, (b) $(BN)_{2n+1}H_{2n+4}$, (c) $(CN)_{2n+1}H_{n+2}$ and (d) $Na_{4n+2}$ (n= 1-5) clusters.

TABLES

**Table 1.** NICS(0) and NICS(1) Values at Different Ring Centers of $C_{4n+2}H_{2n+4}$, $(BN)_{2n+1}H_{2n+4}$, $(BO)_{2n+1}H_{n+2}$, $(CN)_{2n+1}H_{n+2}$ and $Na_{4n+2}$, n=1-5.

| NICS(0) (NICS(1)) | $C_{4n+2}H_{2n+4}$ | $(BN)_{2n+1}H_{2n+4}$ | $(BO)_{2n+1}H_{n+2}$ | $(CN)_{2n+1}H_{n+2}$ | $Na_{4n+2}$ |
|---|---|---|---|---|---|
| A ($A_1$) | -7.91 (-10.14) | -1.41 (-2.63) | -0.30 (-1.97) | -3.96 (-9.65) | -8.90 (-7.79) |
| B ($B_1$) | -8.35 (-10.54) | -0.81 (-1.90) | 7.32 (5.07) | 26.50 (16.67) | -9.50 (-8.26) |
| C ($C_1$) | -7.29 (-9.64) | -0.50 (-1.77) | - | 12.04 (cage) | -9.23 (-8.01) |
| D ($D_1$) | -11.06 (-12.73) | -0.51 (-1.55) | - | 22.94 (cage) | -10.95 (-9.44) |
| E ($E_1$) | -6.30 (-8.65) | -0.51 (-1.67) | - | 0.72 (-2.70) | -8.85 (-7.64) |
| F ($F_1$) | -11.06 (-12.64) | -0.24 (-1.25) | - | 1.12 (-0.38) | -11.24 (-9.68) |
| G ($G_1$) | -5.51 (-7.97) | -0.43 (-1.63) | - | 1.59 (-2.11) | -8.21 (-7.06) |
| H ($H_1$) | -10.54 (-12.19) | -0.33 (-1.30) | - | 4.28 (1.98) | -11.18 (-9.56) |
| I ($I_1$) | -12.14 (-13.46) | -0.11 (-1.13) | - | 2.79 (1.05) | -12.12 (-10.38) |

**Table 2.** Correlation Coefficient ($R^2$) with SD of the Quadratic Behavior of the Ionization Potential (I), Electron Affinity (A), Hardness (η), Softness (S), Chemical Potential (μ), Electrophilicity (ω), Energy (E) and Polarizability (α) as a Function of n of the Acene Analogues.

| | Regression Model: $Z = C_1(SE) + C_2(SE) \times n + C_3(SE) \times n^2$ ; $Z \equiv I, A, \eta, S, \mu, \omega, E, \alpha$; SE=Standard Error | | | |
|---|---|---|---|---|
| | n (1-5) | | | |
| | $C_{4n+2}H_{2n+4}$ | $(BN)_{2n+1}H_{2n+4}$ | $(CN)_{2n+1}H_{n+2}$ | $Na_{4n+2}$ |
| I | $C_1$=8.088(0.121) $C_2$=-1.174(0.092) $C_3$=0.109(0.015) $R^2$=0.9978; SD=0.0562 | $C_1$=8.459(0.137) $C_2$=-0.732(0.104) $C_3$=0.086(0.017) $R^2$=0.9859; SD=0.0638 | $C_1$=9.693(0.400) $C_2$=-2.026(0.305) $C_3$=0.203(0.050) $R^2$=0.9903; SD=0.1865 | $C_1$=3.892(0.038) $C_2$=-0.421(0.029) $C_3$=0.044(0.005) $R^2$=0.9978; SD=0.0176 |
| A | $C_1$=-0.630(0.080) $C_2$=1.188(0.061) $C_3$=-0.104(0.010) $R^2$=0.9992; SD=0.0375 | $C_1$=-0.238(0.017) $C_2$=0.351(0.013) $C_3$=-0.032(0.002) $R^2$=0.9995; SD=0.0078 | $C_1$=1.246(0.522) $C_2$=0.964(0.398) $C_3$=-0.128(0.065) $R^2$=0.8384; SD=0.2433 | $C_1$=1.676(0.018) $C_2$=0.182(0.013) $C_1$=-0.008(0.002) $R^2$=0.9992; SD=0.0082 |
| η | $C_1$=4.359(0.100) $C_2$=-1.181(0.076) $C_3$=0.106(0.012) $R^2$=0.9986; SD=0.0468 | $C_1$=4.348(0.076) $C_2$=-0.542(0.058) $C_3$=0.059(0.010) $R^2$=0.9937; SD=0.0355 | $C_1$=4.224(0.461) $C_2$=-1.495(0.351) $C_3$=0.165(0.057) $R^2$=0.9692; SD=0.2148 | $C_1$=1.108(0.018) $C_2$=-0.302(0.014) $C_3$=0.026(0.002) $R^2$=0.9993; SD=0.0085 |
| 2S | $C_1$=0.207(0.001) $C_2$=0.084(0.001) $C_3$=0.011(0.0001) $R^2$=1.0000; SD=0.0005 | $C_1$=0.222(0.004) $C_2$=0.041(0.003) $C_3$=-0.004(0.001) $R^2$=0.9977; SD=0.0018 | $C_1$=0.154(0.193) $C_2$=0.195(0.147) $C_3$=0.004(0.024) $R^2$=0.9680; SD=0.0900 | $C_1$=0.914(0.027) $C_2$=0.200(0.020) $C_3$=0.085(0.003) $R^2$=0.9999; SD=0.0125 |
| μ | C1=-3.729(0.021) C2=-0.007(0.016) C3=-0.002(0.003) $R^2$=0.9617;SD=0.0098 | $C_1$=-4.111(0.061) $C_2$=0.191(0.046) $C_3$=-0.027(0.008) $R^2$=0.919; SD=0.028 | $C_1$=-5.469(0.062) $C_2$=0.531(0.047) $C_3$=-0.037(0.008) $R^2$=0.9982; SD=0.0290 | $C_1$=-2.784(0.023) $C_2$=0.120(0.018) $C_3$=-0.018(0.003) $R^2$=0.964; SD=0.0107 |
| ω | $C_1$=1.4599(0.031) $C_2$=0.556(0.024) $C_3$=0.097(0.004) $R^2$=0.9999; SD=0.0146 | $C_1$=1.914(0.031) $C_2$=0.102(0.024) $C_3$=-0.002(0.004) $R^2$=0.9946; SD=0.014 | $C_1$=2.714(1.868) $C_2$=1.760(1.423) $C_3$=-0.114(0.233) $R^2$=0.885; SD=0.871 | $C_1$=3.832(0.217) $C_2$=0.130(0.165) $C_3$=0.378(0.027) $R^2$=0.9997; SD=0.1011 |
| E | $C_1$=-78.626(0.002) $C_2$=-153.677(0.002) $C_3$=0.002(0.0003) $R^2$=1.0000; SD=0.0012 | $C_1$=-82.076(0.0002) $C_2$=-186.326(0.0002) $C_3$=$8.10^{-5}$($3.10^{-5}$) $R^2$=1.0000; SD=0.0001 | $C_1$=-94.105(0.017) $C_2$=-186.326(0.013) $C_3$=$1.10^{-04}$(0.00214) $R^2$=1.0000; SD=0.0080 | $C_1$=-324.511(0.079) $C_2$=-649.316(0.060) $C_3$=0.022(0.010) $R^2$=1.0000; SD=0.0369 |
| α | $C_1$=26.574(0.296) $C_2$=33.413(0.226) $C_3$=5.614(0.037) $R^2$=1.0000; SD=0.1382 | $C_1$=22.614(0.530) $C_2$=35.808(0.404) $C_3$=0.688(0.066) $R^2$=1.0000; SD=0.2471 | $C_1$=24.063(13.909) $C_2$=17.005(10.600) $C_3$=10.078(1.733) $R^2$=0.9986; SD=6.4852 | $C_1$=303.695(10.014) $C_2$=260.722(7.631) $C_3$=109.453(1.248) $R^2$=1.0000; SD=4.6691 |

**SYNOPSIS (TOC)**

**Aromaticity in Polyacene Analogues of Inorganic Ring Compounds**

**Pratim Kumar Chattaraj* and Debesh Ranjan Roy**

The aromaticity in the polyacene analogues of several inorganic ring compounds (BN-acenes, CN-acenes, BO-acenes and $Na_6$-acenes) is reported here for the first time. Conceptual density functional theory based reactivity descriptors and the nucleus independent chemical shift (NICS) values are used in this analysis.

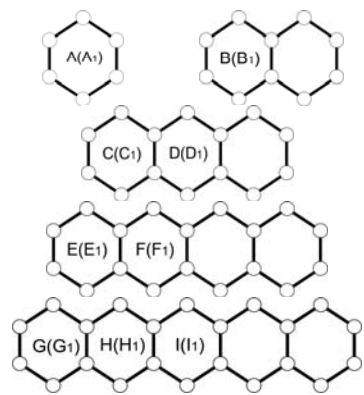